# External Magnetic Fields Enhance Capture of Magnetic Nanoparticles Flowing Through Molded Microfluidic Channels by Ferromagnetic Nanostructures

Reyne Dowling[1] and Mikhail Kostylev[1]

[1]School of Physics, University of Western Australia, Perth, WA, Australia, reyne.dowling@research.uwa.edu.au

**Magnetic nanoparticles (MNPs) have many applications in nanotechnologies. These applications require the particles to be captured and immobilized to allow for their manipulation and sensing. For example, magnetic nanoparticle sensors based on detecting changes to the ferromagnetic resonances of an antidot nanostructure exhibit better performance when the nanoparticles are captured within the antidot inclusions. This study investigates the influence of microfluidics upon the capture of magnetic nanoparticles by four geometries of antidot array nanostructures etched into 30 nm-thick Permalloy films. The nanostructures were exposed to a dispersion of 130 nm magnetic nanoparticle clusters as they passed through PDMS microfluidic channels with a 400 μm circular cross-section fabricated from wire molds. With the microfluidic flow of MNPs, the capture efficiencies - the ratio between the numbers of nanoparticles captured inside of the antidot inclusions to the numbers outside the inclusions - decreased for all four geometries compared to previous results introducing the particles via droplets on the film surface. This indicates that most of the particles were passing over the nanostructures, since there were no significant magnetophoretic forces acting upon the particles. However, when a static magnetic field is applied, the magnetophoretic forces generated by the nanostructure are stronger and the capture efficiencies are significantly higher than those obtained using droplets. In particular, circular antidots demonstrated the highest capture efficiency among the four geometries of almost 83.1% when the magnetic field is parallel to the film plane. In a magnetic field perpendicular to the film, the circle antidots again show the highest capture efficiency of about 77%. These results suggest that the proportion of nanoparticles captured inside the antidot inclusions is highest under a parallel magnetic field. Clearly, the geometry of the nanostructure has a strong influence on the capture of magnetic nanoparticles.**

## I. Introduction and background

Magnetic nanoparticles (MNPs) are indispensable to many nanotechnologies [1-3]. Most MNPs consist of iron-oxide cores less than 100 nm in diameter coated in a protective layer such as dextran. These particles are so small that they display superparamagnetic properties and are usually considered to be single magnetic domains. When magnetized by an external magnetic field, the nanoparticles can be finely manipulated for a number of applications, such as magnetic particle imaging (MPI) and cell sorting [4-7]. Magnetic nanoparticles are also commonly used as catalysts in microscopic chemical reactions in which the particles mix or heat the chemical via magnetic hypothermia[8]. The nanoparticles can be functionalized with specialized coatings for further applications. For instance, some cancer treatments are coated onto magnetic nanoparticles and directed through a patient's body using an external magnetic field [9, 10]. MNPs are particularly useful in biosensing applications such as medical diagnostics and testing water quality [11-17]. Most magnetic biosensors perform assays using magnetic nanoparticles to detect or quantify the presence of a particular protein or virus target in a sample [15, 17, 18]. The assay sample is passed through a suspension of magnetic nanoparticles with coatings that bind, via sandwich assays, the particles to the targets. The bound nanoparticles, which now mark the targets, are then captured and detected by a sensing surface or structure. The sensing area's ability to capture MNPs from the sample is important since it determines how few MNPs are required for detection, which indicates the biosensor's limit of detection (LoD).

Before discussing how magnetic nanoparticles are captured, it is important to understand how MNPs are detected, and what structures are used. MNPs are detected and counted using many techniques, including fluorescent imaging and measurements of magnetoresistance [16, 19-21]. Magnetoresistive sensors observe the nanoparticles indirectly through changes in the Hall voltage passing through thin ferromagnetic multilayer spin-valves. Many of these devices have successfully detected MNPs flowing through microfluidic channels passing over the sensing area [22-28]. In most of these devices, the sensing area is functionalized with biomolecules that bind to and capture the nanoparticles onto the surface of the sensor. Henriksen has shown that fluctuations in the distribution of nanoparticles on a flat magnetoresistive sensor can cause significant fluctuations in the sensor signal [29]. Their results emphasize the importance of understanding, predicting and optimizing the capture of magnetic nanoparticles for improved sensor performance. Chatterjee et al have introduced a new method of nanoparticle detection relying upon exciting ferromagnetic resonance (FMR) in the nanoparticles [30]. In this method, the nanoparticles are detected directly by observing ferromagnetic resonances excited by microwave magnetic fields. An improved method was suggested by Sushruth and Metaxas, who demonstrated that indirect detection using a ferromagnetic antidot nanostructure increased sensitivity and a lowered the LoD [31, 32]. Rather than exciting resonances in the nanoparticles themselves, FMR could be excited in a ferromagnetic nanostructure. When the MNPs are caught inside the nanostructure, they produce noticeable changes in the nanostructure's FMR. The location and number of the nanoparticles caught within this antidot nanostructure may determine the magnitude of the changes that occur in the FMR [33]. Thus, for FMR-based detection of magnetic nanoparticles, controlling where the particles are captured is essential.

A magnetic field is often used to direct the motion of MNPs and collect them in desired locations [7, 26, 34-37]. For example, Cardoso employed a neodymium magnet to separate MNPs from a suspension flowing through a series of microfluidic channels [38]. This principle is also being used to remove microbes in water treatment systems [39]. Magnetophoresis draws the MNPs along the gradient of the applied magnetic field [40, 41]. The magnetophoretic force $\vec{F_m}$ that is exerted onto each magnetic nanoparticle by the magnetic field $\vec{B}$ follows the equation

$$\vec{F_m} = A\,\vec{\nabla}|\vec{B}|^2, \qquad \text{Eq.1}$$

where the constant A is defined by

$$A = \frac{V_{mnp}\,(\chi_{mnp} - \chi_{fluid})}{2\,\mu_0}. \qquad \text{Eq.2}$$

In these equations, $V_{mnp}$ is the volume of the nanoparticle, $\chi_{mnp}$ is the magnetic susceptibility of the nanoparticle, $\chi_{fluid}$ is the magnetic susceptibility of the water and $\mu_0$ is the vacuum permeability.

Many magnetic biosensors also use magnetophoresis to direct and capture the nanoparticles onto the sensing areas[6, 7]. The magnetophoretic forces can be generated by a constant external magnetic field, such as that of a permanent magnet, or by a temporary magnetic field generated by an electromagnet, depending upon the requirements of the device [42, 43]. For instance, Little et al used a constant external magnetic field to generate magnetic fields on the periphery of four spin-valves used for GMR sensing [25]. Alternatively, Kokkinis et al employed a sequence of microscopic conductors to direct MNPs onto a GMR sensor [24]. Clearly, the magnetophoretic forces present in every MNP sensing device will be different. The magnetophoretic forces must be modelled to determine the ideal magnetic fields that must be applied for efficient nanoparticle capturing. This study will focus on the influence of a static external magnetic field upon the capturing of MNPs inside the ferromagnetic antidot nanostructures of Sushruth et al.

In their experiments with FMR-based magnetic nanoparticle detection, Sushruth introduced droplets of MNP suspensions onto the surface of antidot nanostructures, as shown in Figure 1 [32]. The particles were caught on the nanostructures as the droplets evaporated and shrunk. No external magnetic fields were applied, only the weak local fields generated by the ferromagnetic nanostructure itself were capable of influencing how the nanoparticles were caught. In a previous study, we investigated whether an external magnetic field could enhance the capture of MNPs from droplets deposited onto an antidot nanostructure [33]. Micromagnetic simulations performed using MuMax3 indicated that an external field generates magnetophoretic forces that vary across the surface of the nanostructure [44, 45]. Experiments showed that a magnetic field applied parallel to the surface of the nanostructure promotes capture of MNPs inside of the antidots. Therefore, the parallel configuration resulted in a higher capture efficiency. Conversely, a magnetic field applied perpendicular to the surface of the nanostructure promotes capture of MNPs outside of the antidots. Application of a magnetic field in either direction prevented MNP 'coffee rings' from forming around the edges of the droplets [33, 46].

Unlike most biosensors, Sushruth's ferromagnetic detectors did not use microfluidics, requiring samples containing MNPs to be introduced to the devices manually using pipettes. Microfluidics is complementary to biosensing and integration of the two brings many advantages, such as reduced sample volumes, reduced costs, multiplexing and integration with complex microfluidic systems [22, 23, 27]. This system could therefore be improved by introducing MNPs to the nanostructures using microfluidic channels. Multiple studies have been performed on the capture of flowing magnetic nanoparticles. For example, Ezzaier et al demonstrated that the distribution of magnetic nanoparticles captured by an array of magnetized micropillars is highly dependent upon the orientation of an external magnetic field with respect to the flow and nearly independent of the array geometry [47].

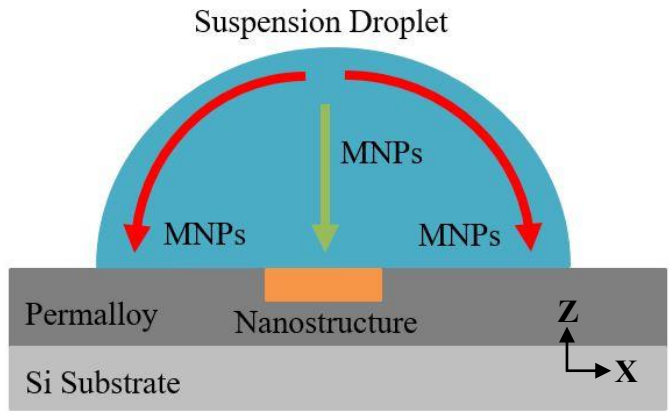

*Figure 1: A droplet containing a suspension of magnetic nanoparticles placed on the surface of a Permalloy film. The droplet is centered over a nanostructure etched into the film. The non-uniformity of the nanostructure generate magnetophoretic forces that attract the MNPs towards the nanostructure. As the droplet dries, outwards capillary flows within the droplet also move the MNPs towards the droplet edge, resulting in a 'coffee ring' of particles on the surface of the film [46].*

However, to date, no studies have investigated the capture of flowing magnetic nanoparticles by ferromagnetic antidot array nanostructures. This study aims to fill this gap in knowledge, as the capture of MNPs is a vital parameter to consider in the design of FMR-based magnetic biosensors. To this end, microfluidic channels were fabricated from poly (dimethyl siloxane) (PDMS) using a soft lithographic technique with wire molds. Three antidot nanostructures were etched into thin Permalloy films via focused ion beam lithography and sealed to the PDMS channels. A suspension of MNPs was passed though the microfluidic channels under parallel and perpendicular external magnetic fields, as shown in Figure 2 below, and a portion of the nanoparticles were captured by the nanostructure. The resulting distributions of nanoparticles caught within the nanostructures were analyzed and compared to the previous results obtained with droplets. The nanostructure capture in microfluidic flow conditions was qualitatively similar to capture with droplets evaporating on the surface of the structure. However, the difference between the parallel and perpendicular magnetic fields was far stronger when capturing from a flow. In particular, the parallel magnetic field captured far more nanoparticles within the antidots than application of a perpendicular magnetic field. The parallel field also demonstrated a higher capture efficiency than for the droplet arrangement, suggesting that microfluidic flow may improve the capture efficiency. These results will be helpful in designing nanostructures that more effectively capture MNPs in future FMR-based magnetic nanoparticle biosensors.

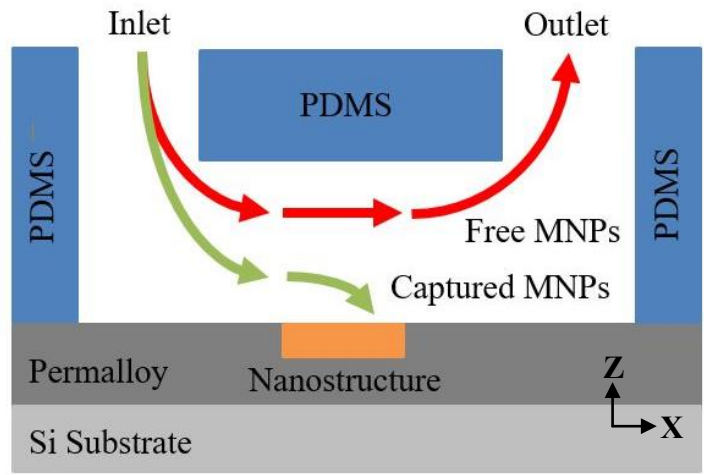

*Figure 2: Magnetic nanoparticles passing through a PDMS microfluidic channel centered over a nanostructure etched into a Permalloy film. Magnetophoretic forces produced by the ferromagnetic nanostructure attract nearby particles towards the nanostructure, which are caught when they come into contact with the surface. Nanoparticles far from the nanostructure are not captured continue into the outlet of the channel.*

## II. MATERIALS AND METHODS

Three ferromagnetic antidot nanostructures were fabricated from Permalloy thin films using focused ion beam lithography. First, a 30 nm layer of Permalloy ($Ni_{80}Fe_{20}$) was coated onto three separate 5 × 5 mm² silicon wafers using magnetron sputtering in an argon atmosphere. The iron and nickel were magnetron sputtered onto the silicon substrates at a temperature of 150 °C and the argon was kept to 6 mTorr. The base pressure in the chamber was lower than $10^{-7}$ Torr. The continuous Permalloy films were then etched using a focused ion beam (FEI Helios FIB-SEM) [48-54]. The beam current was set to 80 pA and the accelerating voltage was 30 kV. The beam etched 15 × 15 square arrays of circular antidots with 400nm diameter and 600nm separation between the centers of the antidots, as shown in Figure 3. To explore the statistics of MNP capture in the antidots, the etching was repeated to obtain 20 separate circular antidot arrays on each of the three films. To investigate the influence of the array geometry on nanoparticle capture, the ion beam also etched 20 arrays of circular dots (the inverse of an antidot nanostructure), square antidots and square dots. In total, each Permalloy film contained 80 separate nanostructures.

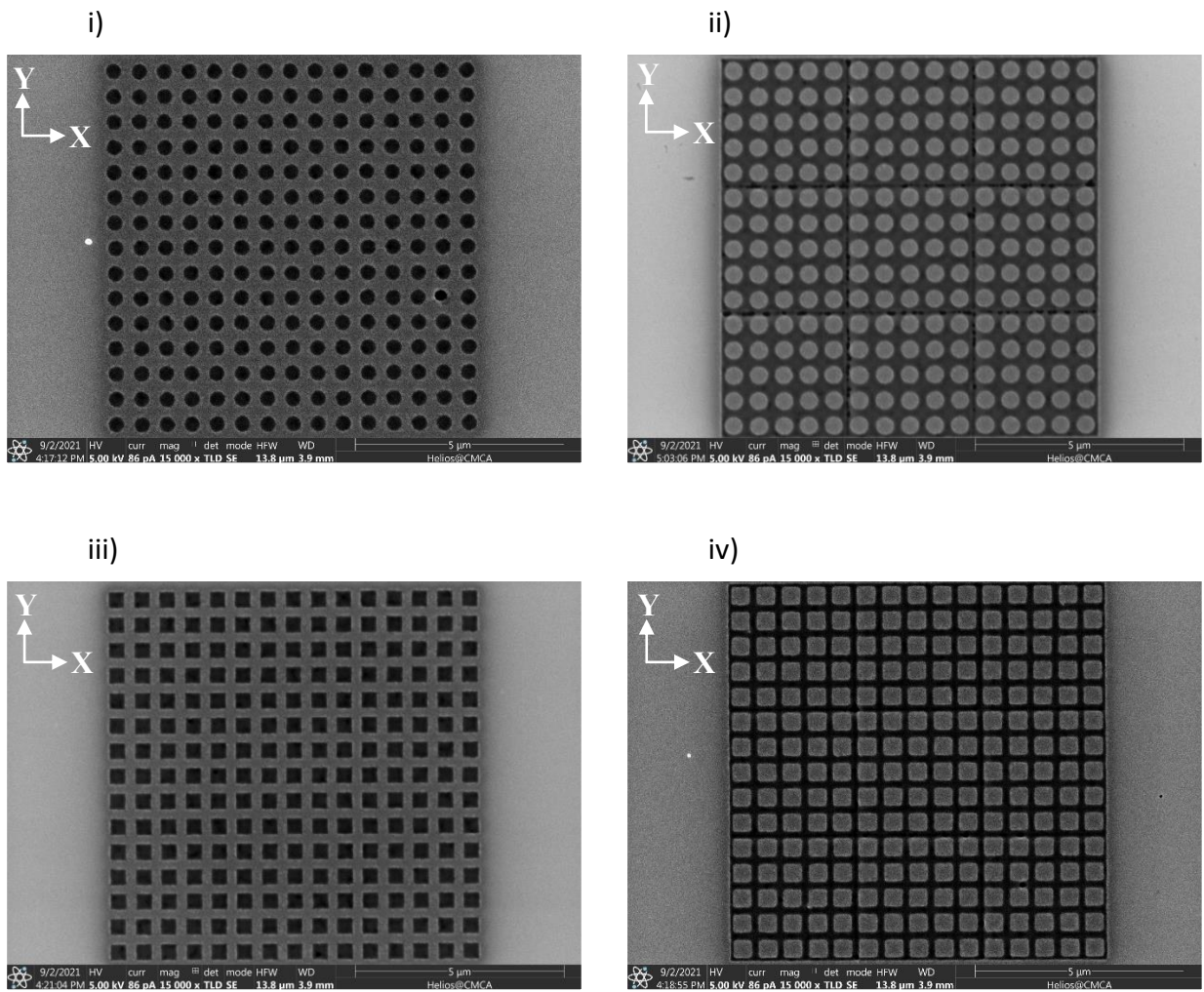

*Figure 3: Images of the four geometries of nanostructures etched into 30-nm thick Permalloy films using a focused ion beam. These images were obtained before introduction of any MNPs. The images show the following: i) an array of circular antidots with 400 nm diameter, ii) an array of circular dots with 400 nm diameter, iii) an array of square antidots with sides of length 400 nm, iv) an array of square dots with sides of length 400 nm. In all four geometries, the centers of the antidots and dots are separated by 600 nm in both direction.*

Microfluidic channels of dimensions 0.4 × 0.4 × 10 mm were fabricated using the soft lithography process shown in Figure 4 [55-57]. Using this method, one can entirely avoid the relatively expensive and complex processes associated with hard lithography. The channel molds were crafted from 10 mm-long sections of 0.4 mm-diameter wire bent into a U shape with a flattened bottom. The molds were then secured onto double-sided tape placed onto the base of a petri dish. A mixture of the PDMS elastomer base (Sylgard® 184 silicone elastomer) and curing agent were poured into the molds in a ratio of 10:1 by volume and left to set for 24 hours in ambient laboratory conditions. String was used to keep the molds still as the PDMS mixture was poured into the petri dish and set. Once the PDMS film was set, the film was removed from the petri dish using a scalpel and the wire molds were removed from the PDMS film using tweezers. Luer stubs (Instech Laboratories, Inc.) were inserted into the ends of the channels, which act as inlets and outlets, for connecting the external tubing. Incorporating the inlet and outlet into the channel mold removes the need for punching holes into the PDMS, which could damage the microfluidic channel. The PDMS films were then pressed onto the surfaces of the Permalloy films, creating a water-tight seal between the two. An optical microscope was used to position the channels so that the nanostructures were located at the centers of the microchannels. Since relatively complex molds can be produced in only a few minutes, this process is particularly useful for fabricating prototype microfluidic devices at low cost. The mold could be 3D printed for applications requiring finer and more uniform microfluidic channels.

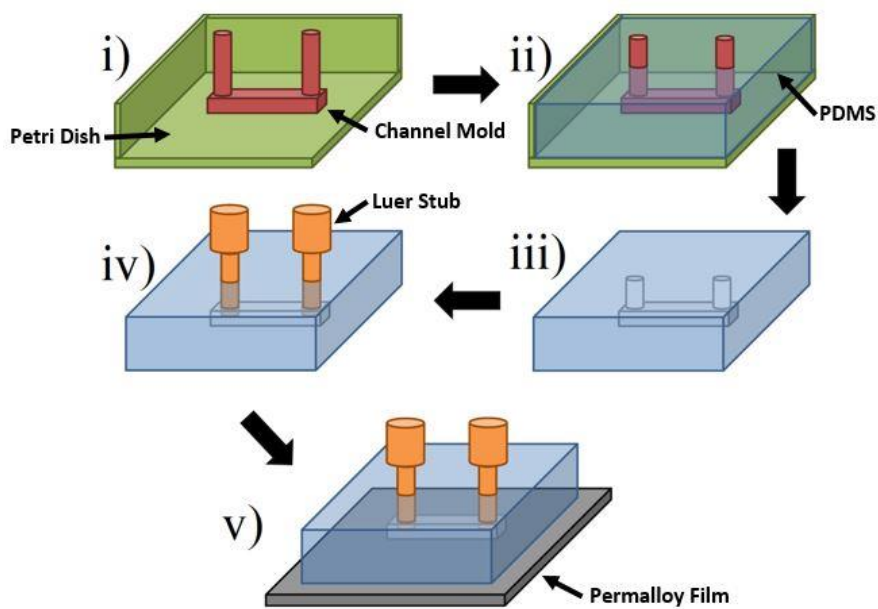

*Figure 4: The process used to fabricate microfluidic channels: i) a mold is created using 0.4 mm-diameter wire (red) and placed inside a container such as a petri dish (green), ii) PDMS elastomer (blue) is poured into the mold, iii) the PDMS sets over 24 hours and is removed from the container. The wire mold is removed from the PDMS film using tweezers, iv) luer stubs (orange) are inserted into the slots at both ends of the channel to form inlet and outlet ports, v) The PDMS film is pressed against the surface of a silicon wafer or nanostructured Permalloy film to seal the microfluidic channel closed.*

The inlets of the microchannels were connected to syringes slotted onto a syringe pump (Harvard Apparatus Pump 11 Elite). Each syringe was filled with 3 mL of 0.1 mg/mL suspensions of dextran-coated, 130 nm-diameter iron-oxide magnetic nanoparticle clusters (Dynamag®-D, micromod Partikeltechnologie GmbH). The concentration of the nanoparticle suspension was originally 25 g/mL, which was diluted to 0.1 mg/mL using distilled water. This concentration was chosen to prevent the nanostructures from being completely covered by nanoparticles and to avoid the formation of MNP agglomerates and chains that occur in dense MNP suspensions under an applied magnetic field [36, 58-61]. The channel outlets were connected via tubing to glass beakers serving as reservoirs.

One of the devices was then placed between two neodymium magnets, which produced a uniform magnetic field of 1.38 kOe across the channel and parallel to the plane of the Permalloy film, as shown in Figures 5 and 6. Another device was placed between two cylindrical magnets producing a 3.5 kOe magnetic field perpendicular to the plane of the film, as shown in Figures 5 and 7. The third device was not placed inside an external magnetic field. The syringe pump was then activated and infused the MNPs through each of the microfluidic channels at a rate of 1 mL/min. Afterwards, 3 mL of distilled water was pumped through each channel to flush the remaining MNPs out of channels before they could settle and another 3 mL of air was pumped through the channels to dry them. Once the infusions were complete, the PDMS films were peeled off of the Permalloy films and the nanostructures were imaged using a scanning electron microscope (FEI Verios XHR SEM).

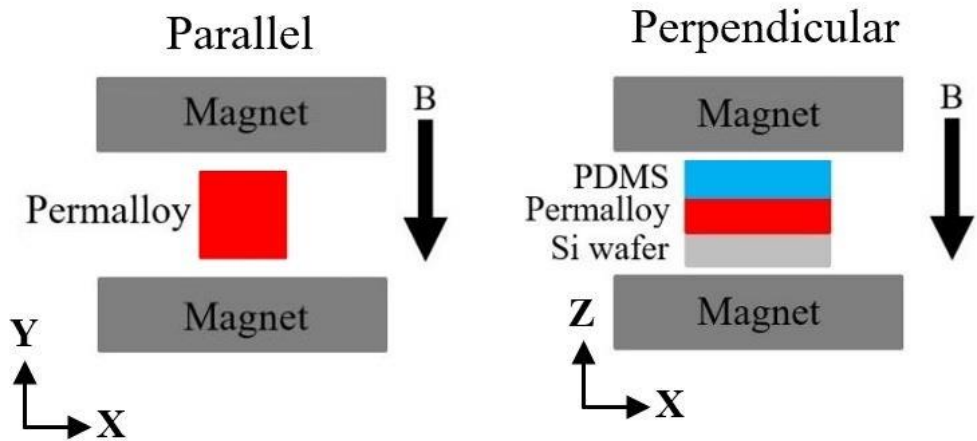

*Figure 5: The two configurations of neodymium permanent magnets producing a static magnetic field either parallel or perpendicular to the Permalloy films. For a parallel magnetic field, the Permalloy film is placed equidistant from two cylindrical magnets, where the field has a strength of 1.38 kOe and is directed along the y axis. For a perpendicular magnetic field, the Permalloy film is equidistant and parallel to two cylindrical magnets, where the magnetic field has a strength of 3.5 kOe and is directed along the z axis.*

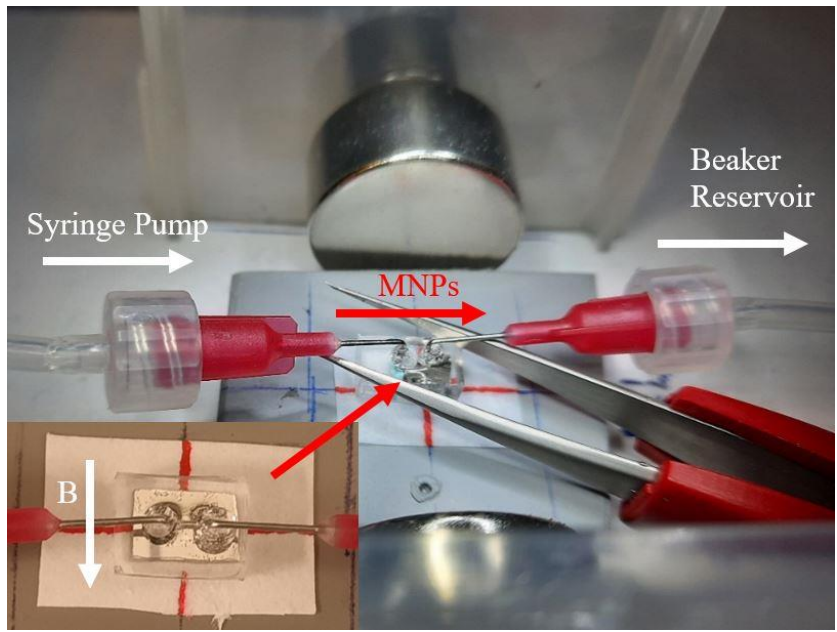

*Figure 6: A photograph of the first experimental apparatus. A syringe pump forces a dispersion of MNPs through a microfluidic channel pressed against the surface of a Permalloy thin film. Two cylindrical permanent magnets produce a magnetic field B directed parallel to the plane of the film, along the y axis, as shown in Figure 5. As the dispersion passes through the microchannel, some of the MNPs are attracted towards the surface of the film by magnetophoretic forces and are captured by arrays of antidots and dots hollowed into the surface. The remaining nanoparticles exit the channel and are collected into a beaker. The inset shows the microfluidic device in further detail. A pair of tweezers was used to press the PDMS microfluidic film onto the Permalloy film, ensuring the seal between the two is not broken and the microfluidic device remains parallel to the magnetic field.*

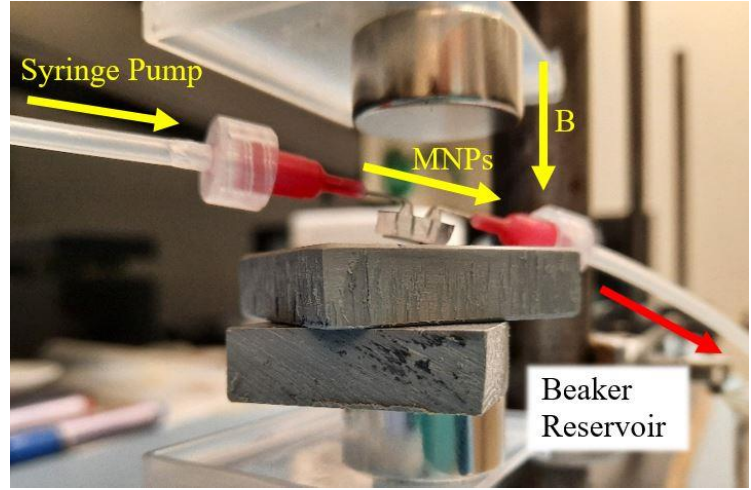

*Figure 7: A photograph of the second experimental apparatus. A syringe pump forces a dispersion of MNPs through a microfluidic channel pressed against the surface of a Permalloy thin film. Two cylindrical permanent magnets produce a magnetic field B directed perpendicular to the plane of the film, along the z axis, as shown in Figure 5. As the dispersion passes through the microchannel, some of the MNPs are attracted towards the surface of the film by magnetophoretic forces and are captured by arrays of antidots and dots hollowed into the surface. The remaining nanoparticles exit the channel and are collected into a beaker. The inset shows the microchannel in further detail. A pair of tweezers was used to press the PDMS microfluidic film onto the Permalloy film, ensuring the seal between the two is not broken and the microfluidic device remains perpendicular to the magnetic field.*

In analyzing the microscope images, the nanoparticles were counted and sorted as captured inside or outside of each nanostructure using the criteria depicted in Figure 8. Particles trapped 'inside' of the nanostructure are caught on the internal surfaces of the nanostructure. Conversely, particles trapped 'outside' of the nanostructure are caught on the outer surface of the structure. The counting results were then used to calculate the average number of nanoparticles captured inside of each nanostructure geometry, $N_{IN}$, and the average number of nanoparticle caught outside of each nanostructure geometry, $N_{OUT}$. Nanostructures with dot geometries have greater volumes for capturing magnetic nanoparticles than their corresponding antidot geometries. This difference between the volume of the material that was removed and the volume that remains is accounted for by scaling the average numbers of particles inside $N_{IN}$ and particles outside $N_{OUT}$ by the scale factors $S_{IN}$ and $S_{OUT}$ respectively, where

$$S_{IN} = \frac{volume\ remaining}{volume\ removed} \qquad \text{Eq. 3}$$

and

$$S_{OUT} = \frac{volume\ removed}{volume\ remaining}. \qquad \text{Eq. 4}$$

Finally, the capture efficiency of the arrays was calculated using Equation 5 below. This number is indicative of each nanostructure geometry's ability to capture MNPs internally and enables direct comparison of the four geometries. The capture efficiency estimates the proportion of MNPs that are or could be captured inside of each nanostructure geometry in comparison to the total number of nanoparticles in the suspension:

$$Capture\ Efficiency = \frac{S_{IN}\ N_{IN}}{S_{IN}\ N_{IN} + S_{OUT}\ N_{OUT}} \qquad \text{Eq. 5}$$

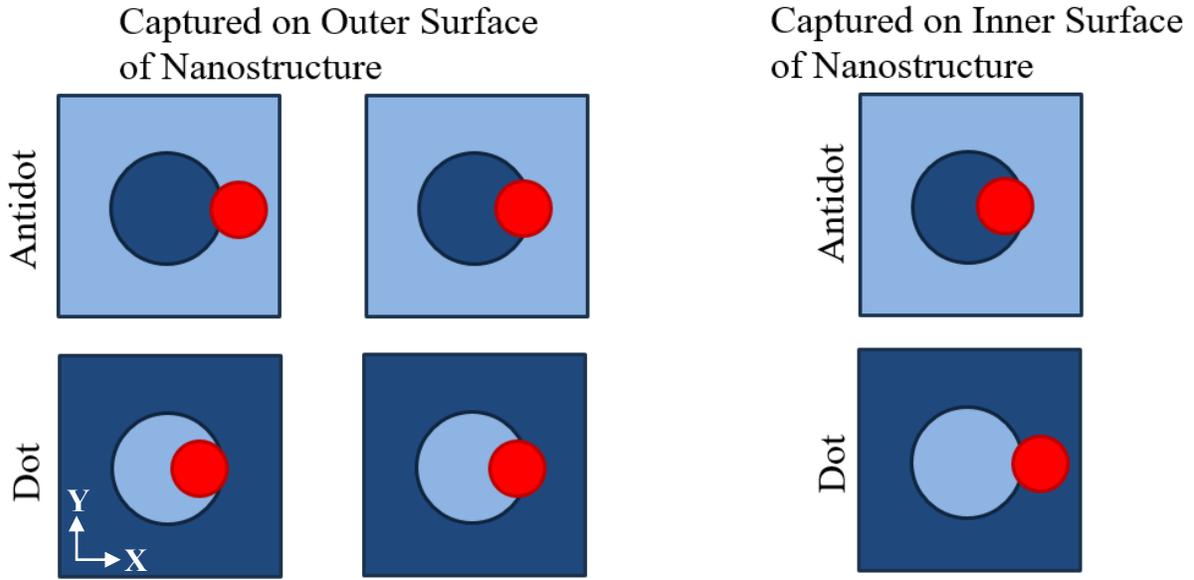

*Figure 8: Diagrams depicting the position of a magnetic nanoparticle (red) with respect to an antidot or dot and indicating whether these nanoparticles are considered to have been captured on the outer or inner surface of the nanostructure. Lighter blue regions represent areas containing Permalloy while darker regions represent empty spaces in which Permalloy has been removed from the film. For a nanoparticle to be considered as caught inside of the nanostructure, most of the nanoparticle must be located in one of the darker regions of the nanostructure. Otherwise, the particle is counted as captured on the outside of the nanostructure.*

## III. Results and Discussion

### A. Capturing Nanoparticles with No External Magnetic Field

After infusion of magnetic nanoparticles, the distributions of particles caught by each of the 80 nanostructures were imaged by scanning electron microscopy. The nanoparticles were then counted and the results were analyzed according to the method outlined in the previous section. Figure 9 shows a typical distribution of magnetic nanoparticles caught by each of the four nanostructure geometries. The distributions are relatively uniform across the antidot and dot arrays, with particles being caught both inside and outside of the antidots. The nanoparticle distributions obtained from MNPs caught from a flow are qualitatively similar to those obtained when the MNPs were deposited onto the nanostructures in droplets. This is reflected in the total and average number of particles counted for each of the four nanostructure geometries, which is tabulated in Table I below. Table I also includes the capture efficiencies obtained from droplet depositions performed in a previous study. As with the distributions obtained from droplets onto the surface of the nanostructures, the antidot geometries show stronger capture efficiencies than the dot geometries. In addition, the circular antidot nanostructures again showed the strongest capture efficiency of 62%, which is lower than the 70.8% obtained from deposition by droplets.

However, the three geometries show significantly lower capture efficiencies than those obtained from droplet depositions. In particular, the efficiency of the square antidot geometry fell from 55.9% to only 29.6%. This suggests that depositing the nanoparticles via a microfluidic flow decreases the ratio of nanoparticles caught inside the nanostructures rather than increasing it. Additionally, the circular antidots captured only 13 MNPs across 20 nanostructures. Given that the magnetic field produced by the ferromagnetic nanostructure at remanence is weak, it is highly likely that most of the nanoparticles in the suspension are simply flowing past the nanostructures. Only the few nanoparticles that travel close to the surface of the nanostructure experience magnetophoretic forces strong enough to attract them towards and onto the nanostructures. Simplified depictions of the flows of magnetic nanoparticles through a microfluidic channel are depicted in Figure 2. Reducing the dimensions of the microfluidic channel would focus the nanoparticles closer to the nanostructures, which may improve the capture efficiencies.

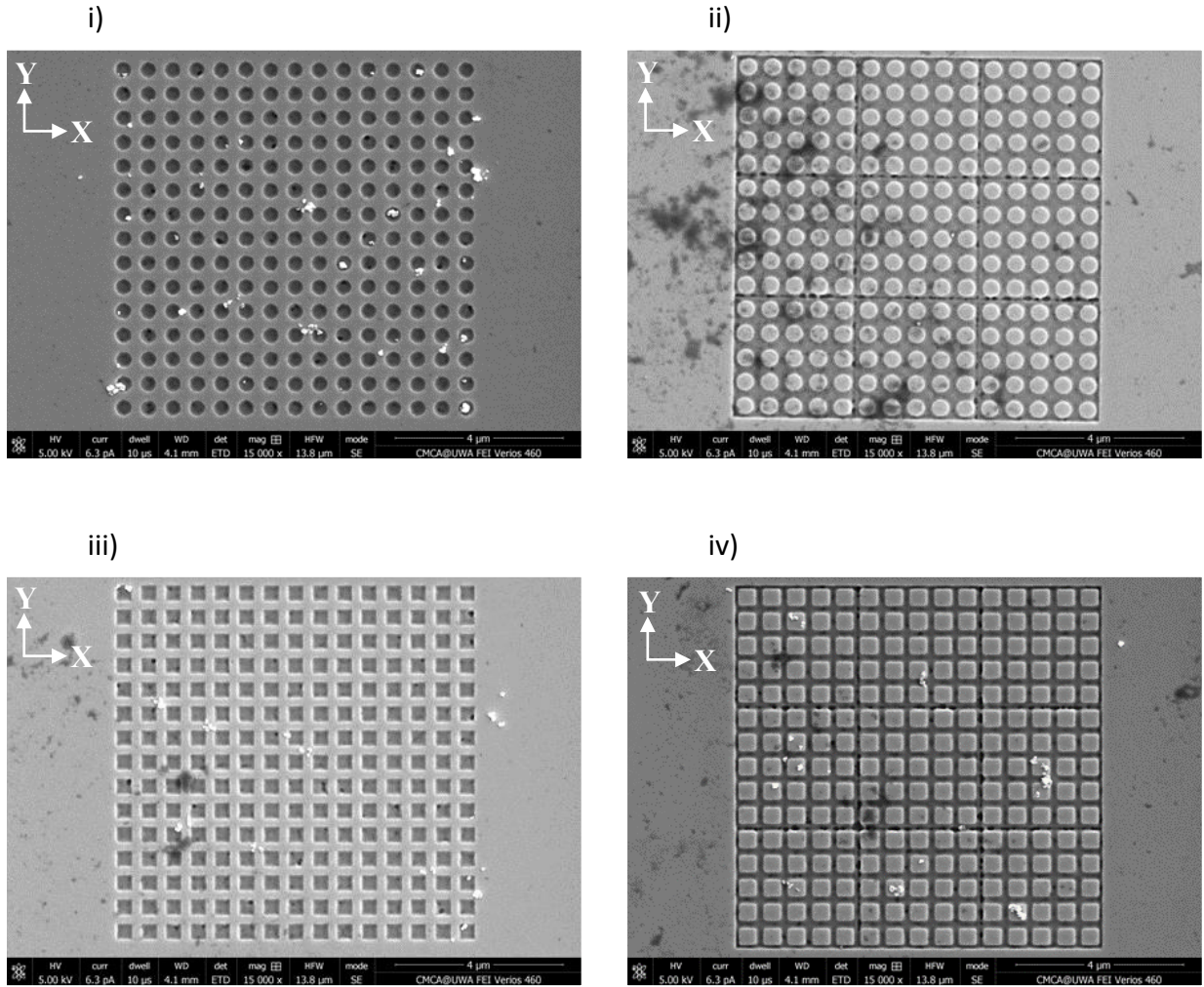

*Figure 9: Images obtained using scanning electron microscopy of the typical distributions of MNPs caught on the surfaces of four different ferromagnetic arrays etched into a thin Permalloy film. The MNPs were introduced to these arrays from a microfluidic flow without applying any static magnetic fields. The geometries of the four nanostructures are: i) circle antidots, ii) circle dots, iii) square antidots, and iv) square dots.*

TABLE I
MAGNETIC NANOPARTICLES CAUGHT WITHOUT EXTERNAL MAGNETIC FIELD

| Nanostructure Geometry | Total Inside | Total Outside | Average Inside | Average Outside | Flow Capture Efficiency (%) | Droplet Capture Efficiency (%) |
|---|---|---|---|---|---|---|
| Circle Antidots | 152 | 321 | 12.88 | 7.82 | 62.2 | 70.8 |
| Circle Dots | 7 | 6 | 0.19 | 0.62 | 23.5 | 21.3 |
| Square Antidots | 50 | 195 | 3.13 | 7.43 | 29.6 | 55.9 |
| Square Dots | 191 | 354 | 7.28 | 21.07 | 25.7 | 35.7 |

The total number of magnetic nanoparticles caught inside and outside of 15 × 15 arrays of antidot inclusions etched into a single Permalloy film. The MNPs were introduced to the ferromagnetic nanostructures via a microfluidic flow without applying a static magnetic. The total and average counts were obtained from 20 samples for each geometry of nanostructure. These values have been weighted according to Eq. 3 and Eq. 4 to account for the differences in the total volumes of the inclusions between each geometry. The capture efficiency for each nanostructure geometry was calculated using Eq. 5.

## B. Capturing Nanoparticles with an External Magnetic Field Parallel to the Film

Figure 10 shows four SEM images representative of the distributions of MNPs captured by each of the four geometries of nanostructures under an external magnetic field directed parallel to the films. The total count of MNPs captured inside and outside of each of the four geometries have been collected in Table II, along with the capture efficiencies of each geometry. The images and capture efficiencies indicate that almost half, or more, of the nanoparticles are caught within the inclusions of the nanostructure. In particular, the circle dots exhibited the highest capture efficiency of 83.1% - more than the highest efficiency obtained, also in circle dots, of 70.5% when the MNPs are introduced via droplets. The lowest capture efficiency of all four geometries was again achieved by the circle dots. This capture efficiency was 43.9% - only 0.2% lower than the 44.1% observed when MNPs are introduced via droplets. These results suggest that the integration of microfluidics rather than droplets under a magnetic field applied parallel to the plane of the nanostructured film generally increases the capture efficiency. This may be due to the microfluidic flow removing any nanoparticles that are weakly bound to the Permalloy nanostructure. Naturally, the particles caught on the outer surface of the nanostructure are easier to wash away than those nanoparticles caught inside of the nanostructures.

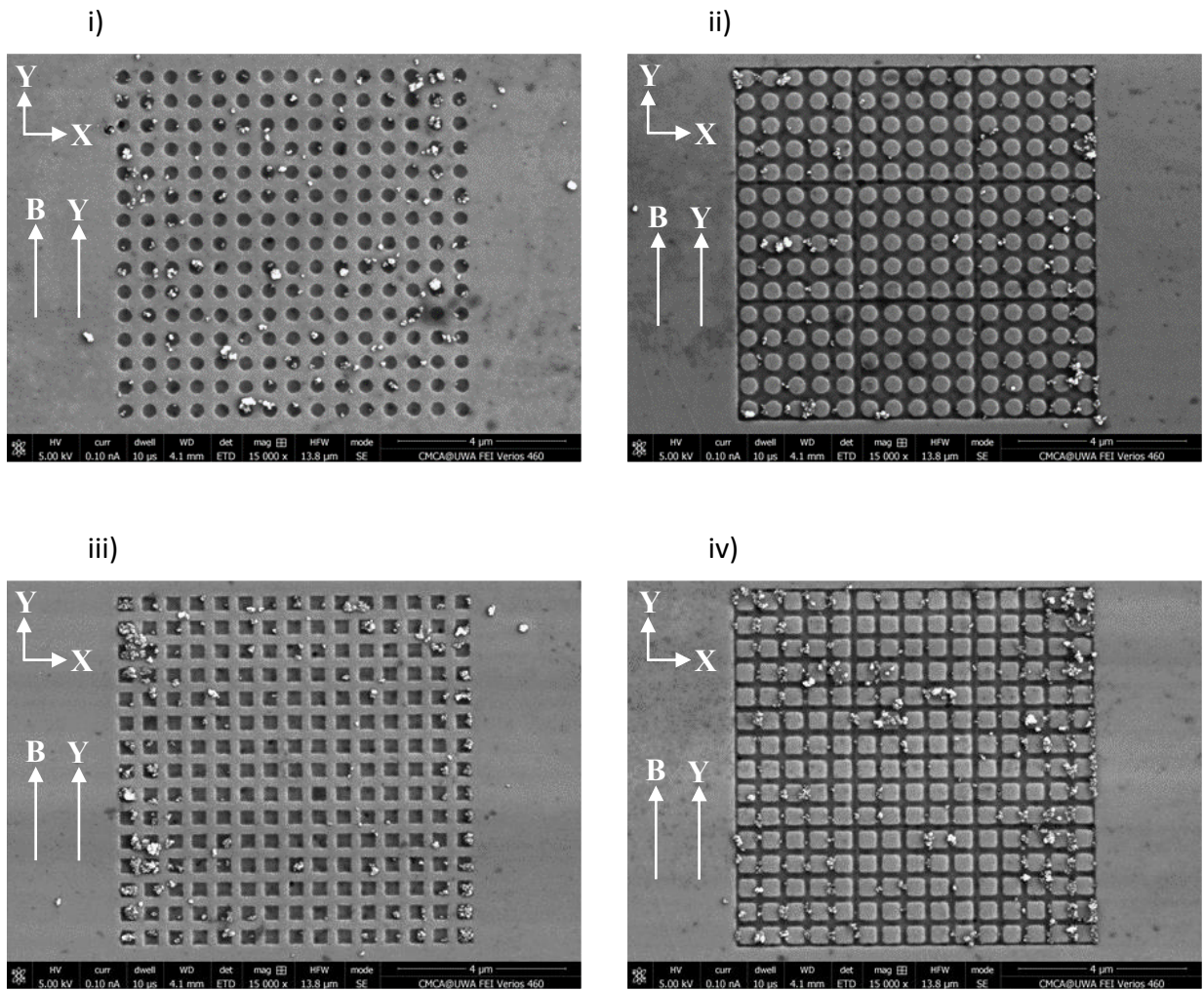

*Figure 10: Images obtained using scanning electron microscopy of the typical distributions of MNPs caught on the surfaces of four different ferromagnetic arrays etched into a thin Permalloy film. The MNPs were introduced to these arrays from a microfluidic flow under a static magnetic field applied parallel to the plane of the film, along the y axis. The geometries of the four nanostructures are: i) circle antidots, ii) circle dots, iii) square antidots, and iv) square dots.*

TABLE II
MAGNETIC NANOPARTICLES CAUGHT WITH EXTERNAL MAGNETIC FIELD PARALLEL TO FILM

| Nanostructure Geometry | Total Inside | Total Outside | Average Inside | Average Outside | Flow Capture Efficiency (%) | Droplet Capture Efficiency (%) |
|---|---|---|---|---|---|---|
| Circle Antidots | 474 | 389 | 46.52 | 9.48 | 83.1 | 70.5 |
| Circle Dots | 363 | 120 | 9.73 | 12.43 | 43.9 | 44.1 |
| Square Antidots | 1095 | 493 | 72.04 | 18.78 | 79.3 | 63.9 |
| Square Dots | 1204 | 421 | 48.16 | 25.06 | 65.8 | 47.7 |

The total number of magnetic nanoparticles caught inside and outside of 15 × 15 arrays of antidot inclusions etched into a single Permalloy film. The MNPs were introduced to the ferromagnetic nanostructures via a microfluidic flow under a static magnetic field directed parallel to the plane of the Permalloy film. The total and average counts were obtained from 20 samples for each geometry of nanostructure. These values have been weighted according to Eq. 3 and Eq. 4 to account for the differences in the total volumes of the inclusions between each geometry. The capture efficiency for each nanostructure geometry was calculated using Eq. 5.

Application of a parallel magnetic field has significantly increased the ratio of MNPs captured inside of the nanostructures. The greatest rise in capture efficiency was observed in the antidot nanostructures. For example, upon addition of a static magnetic field parallel to the plane of the Permalloy film, the capture efficiency of square antidot nanostructures has increased from 29.6% to 79.3%. This dramatic change in the efficiencies is likely due to the parallel magnetic field generating attractive magnetophoretic forces over the antidots [33]. In addition, the total number of MNPs caught by all four geometries is far greater than without a magnetic field. This is likely due to the parallel magnetic field strengthening the magnetization of the nanostructure. This produces stronger local magnetic fields that penetrate further into the microfluidic channel. This results in larger magnetophoretic forces, which act to direct more MNPs towards the nanostructure.

C. *Capturing Nanoparticles with an External Magnetic Field Perpendicular to the Film*

Images of the distributions of MNPs captured from the microfluidic flow in a perpendicular magnetic field are displayed in Figure 11. The results of counting these nanoparticles are summarized in Table III. When introducing the MNPs to the sensor using droplets, the capture efficiencies were found to be particularly low when a magnetic field was applied perpendicular to the plane of the Permalloy film. For this configuration, the highest capture efficiency was observed in the square antidot nanostructures at only 28.8%. In addition, the circle dot geometry displayed the lowest efficiency of only 7.5%, significantly lower than the other three geometries. These results suggest that application of a perpendicular external magnetic field is undesirable for applications such as FMR-based MNP sensors which rely on nanoparticles being caught within ferromagnetic nanostructures. However, when the nanoparticles are introduced via a microfluidic channel rather than in droplets, the capture efficiencies are noticeably higher. With microfluidics, the highest capture efficiency of 77% was obtained from the circle antidot geometry. This value is considerably larger than the value obtained when introducing the MNPs via droplet and is comparable to the largest efficiency in circle antidots of 83.1% obtained with microfluidics under application of a parallel magnetic field.

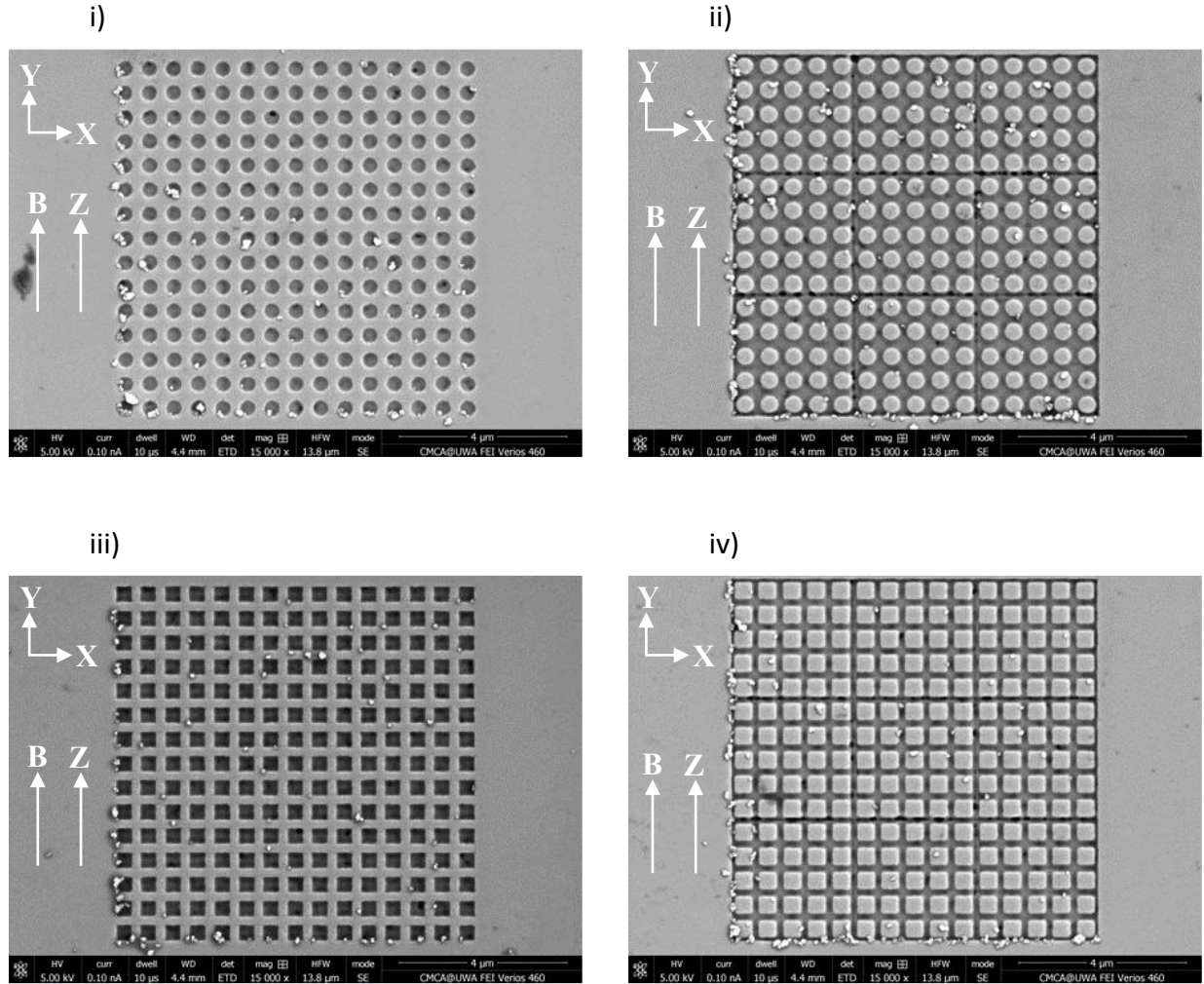

*Figure 11: Images obtained using scanning electron microscopy of the typical distributions of MNPs caught on the surfaces of four different ferromagnetic arrays etched into a thin Permalloy film. The MNPs were introduced to these arrays from a microfluidic flow under a static magnetic field applied perpendicular to the plane of the film along the z axis. The geometries of the four nanostructures are: i) circle antidots, ii) circle dots, iii) square antidots, and iv) square dots.*

TABLE III
MAGNETIC NANOPARTICLES CAUGHT WITH EXTERNAL MAGNETIC FIELD PERPENDICULAR TO FILM

| Nanostructure Geometry | Total Inside | Total Outside | Average Inside | Average Outside | Flow Capture Efficiency (%) | Droplet Capture Efficiency (%) |
|---|---|---|---|---|---|---|
| Circle Antidots | 254 | 277 | 22.56 | 6.75 | 77.0 | 24.1 |
| Circle Dots | 323 | 400 | 7.87 | 41.44 | 16.0 | 7.5 |
| Square Antidots | 173 | 268 | 10.3 | 10.21 | 50.2 | 28.8 |
| Square Dots | 206 | 295 | 7.85 | 17.56 | 30.9 | 20.4 |

The total number of magnetic nanoparticles caught inside and outside of 15 × 15 arrays of antidot inclusions etched into a single Permalloy film. The MNPs were introduced to the ferromagnetic nanostructures via a microfluidic flow under a static magnetic field directed perpendicular to the plane of the Permalloy film. The total and average counts were obtained from 20 samples for each geometry of nanostructure. These values have been weighted according to Eq. 3 and Eq. 4 to account for the differences in the total volumes of the inclusions between each geometry. The capture efficiency for each nanostructure geometry was calculated using Eq. 5.

However, the other three geometries exhibit much lower capture efficiencies under a perpendicular magnetic field than in a parallel field. Far fewer MNPs were captured from the microfluidic flow under a perpendicular magnetic field. This was expected, since simulations have shown that a perpendicular magnetic field results in repulsive magnetophoretic forces above antidots [33]. Still, application of a perpendicular magnetic field still resulted in more nanoparticles being captured from the microfluidic flow than in the arrangement without a magnetic field. This suggests that application of static magnetic fields in any orientation will generally increase the total number of magnetic nanoparticles captured by the ferromagnetic nanostructure. Applying a magnetic field increases the magnetization of the structure, increasing the magnitude of the magnetophoretic forces generated by the structure.

One can also observe many MNPs are caught along two edges of the nanostructure, but only when the static magnetic field is perpendicular to the film. This was also observed in previous experiments in which the MNPs were introduced via droplets. Simulations suggest that this is due to the static magnetic field being too weak to completely saturate the ferromagnetic nanostructure, causing regions of attractive magnetophoretic forces along two edges of the nanostructure.

## IV. Conclusions

Four geometries of ferromagnetic antidot array nanostructures were integrated with microfluidic channels to observe the distributions of magnetic nanoparticles captured from the microfluidic flow by the nanostructures. These flow distributions were compared to previous distributions obtained when magnetic nanoparticles are introduced onto the surface of the films in droplets. The microfluidic channels were fabricated simply using a wire mold, avoiding the need for any relatively expensive hard-lithographic processes. When integrated onto nanostructures etched into Permalloy thin films, the ratio of particles captured within the inclusions of the structures increased for all four nanostructure geometries. The flow of water appears to remove MNPs bound weakly to the nanostructure surface, leaving particles mostly inside the antidot inclusions. Increasing the number of particles caught inside the antidot inclusions may be useful for applications such as magnetic nanoparticle sensors, collectors and manipulators.

Nanoparticle distributions from microfluidic flows were also obtained under external fields directed parallel and perpendicular to the planes of the Permalloy films. The results across the two arrangements of the external magnetic field and without a field indicate that circle antidot nanostructures perform far better at internally capturing MNPs from a microfluidic flow, even under a perpendicular magnetic field. In addition, the capture efficiencies for antidot geometries are consistently higher than for the dot geometries. The geometry of the nanostructure clearly influences the magnetophoresis and capture of MNPs. The geometry of the ferromagnetic nanostructure should be considered for any systems designed to capture, manipulate or sense magnetic nanoparticles.

## Acknowledgements

R. D. acknowledges a RTP Stipend received from the University of Western Australia. The authors also thank the staff at the Centre for Microscopy, Characterisation and Analysis (CMCA) for access to and assistance with scanning electron microscopy.